\documentclass[preprint,eqsecnum,aps,nofootinbib]{revtex4}
\usepackage{amsfonts,amsmath,amssymb,amsthm}
\usepackage{latexsym}
\usepackage{bbm,bm}
\usepackage{graphicx}

\newcommand{\beq}{\begin{equation}}
\newcommand{\eeq}{\end{equation}}
\newcommand{\beqs}{\begin{eqnarray}}
\newcommand{\eeqs}{\end{eqnarray}}

\begin{document}

\title{Generalized uncertainty principle and $D$-dimensional quantum mechanics}

\author{DaeKil Park$^{1,2}$\footnote{dkpark@kyungnam.ac.kr}}

\affiliation{$^1$Department of Electronic Engineering, Kyungnam University, Changwon
                 631-701, Korea    \\
             $^2$Department of Physics, Kyungnam University, Changwon
                  631-701, Korea    
                      }

\begin{abstract}
The non-relativistic quantum mechanics with a generalized uncertainty principle (GUP) is examined in $D$-dimensional free particle and harmonic oscillator systems.
The Feynman propagators for these systems are exactly derived within the first order of the GUP parameter.

\end{abstract}

\maketitle

\section{Introduction}

The existence of a minimal length seems to be a model-independent feature of quantum gravity\cite{townsend76,amati89,garay94}. It appears as various different expressions in loop quantum gravity\cite{rovelli98,carlip01},
string theory\cite{konishi90,kato90}, path-integral quantum gravity\cite{padmanabhan85,padmanabhan87,greensite91}, and black hole physics\cite{maggiore93}. From the aspect of quantum mechanics the existence of a minimal 
length results in the modification of the Heisenberg uncertainty principle (HUP)\cite{uncertainty,robertson1929} $\Delta P \Delta Q \geq \frac{\hbar}{2}$, because $\Delta Q$ should be larger than the minimal length.
Various modification of HUP, called the generalized uncertainty principle (GUP), were suggested in Ref. \cite{kempf93,kempf94}.

The purpose of the paper is to examine the $D$-dimensional non-relativistic quantum mechanics when the GUP has a following specific form
\begin{equation}
\label{GUP-d-1}
\Delta P_i \Delta Q_i \geq \frac{\hbar}{2} \left[ 1 + \alpha \left(\Delta {\bf P}^2 + \langle {\bf P} \rangle^2 \right) + 2 \alpha \left( \Delta P_i^2 + \langle P_i \rangle^2 \right) \right]
\hspace{1.0cm}  (i = 1, 2, \cdots, D)
\end{equation}
where $\alpha$ is a GUP parameter, which has a dimension $(\mbox{momentum})^{-2}$. Using $\Delta A \Delta B \geq \frac{1}{2} | \langle [A, B] \rangle |$, Eq. (\ref{GUP-d-1}) induces the 
modification of the commutation relation as 
\begin{eqnarray}
\label{GUP-d-2}
&& \left[ Q_i, P_j \right] = i \hbar \left( \delta_{ij} + \alpha \delta_{ij} {\bf P}^2 + 2 \alpha P_i P_j  \right)    \\    \nonumber
&& \hspace{1.0cm} \left[ Q_i, Q_j \right] = \left[P_i, P_j \right] = 0.
\end{eqnarray}

\begin{figure}[ht!]
\begin{center}
\includegraphics[height=6.0cm]{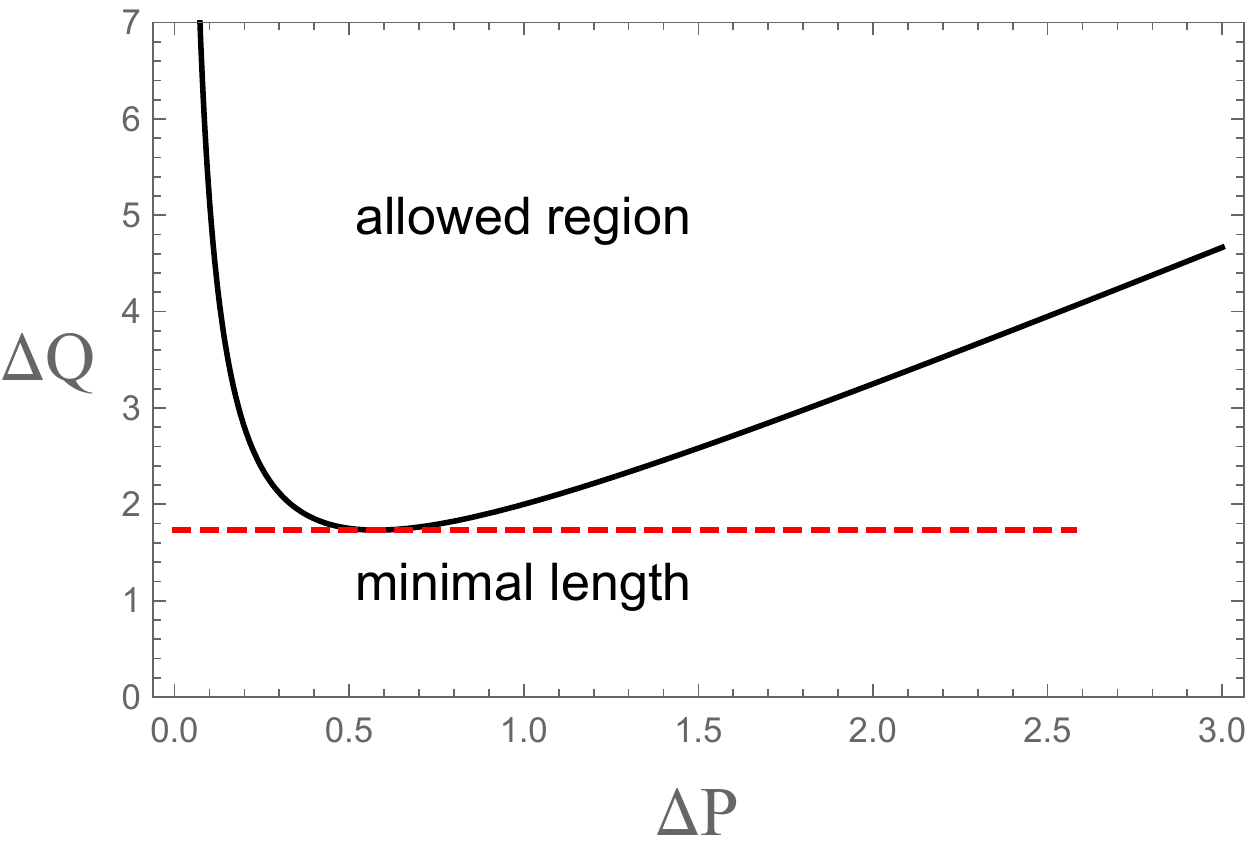} 

\caption[fig1]{(Color online) The minimal length and allowed region of one-dimensional GUP (\ref{GUP-d-3}) when $\hbar = \alpha = 1$. 
 }
\end{center}
\end{figure}

The existence of the minimal length is easily shown at $D = 1$. In this case  Eq. (\ref{GUP-d-1}) is expressed as 
\begin{equation}
\label{GUP-d-3}
\Delta P \Delta Q \geq \frac{\hbar}{2} \left( 1 + 3 \alpha \Delta P^2 \right)
\end{equation}
if $\langle P \rangle = 0$. Then, the equality of Eq. (\ref{GUP-d-3}) yields 
\begin{equation}
\label{minimal-length}
\Delta Q^2 \geq \Delta Q_{min}^2 = 3 \alpha \hbar^2.
\end{equation}
In Fig. 1 the allowed region and minimal length of Eq. (\ref{GUP-d-3}) is plotted when $\hbar = \alpha = 1$. 

If $\alpha$ is small, Eq. (\ref{GUP-d-2}) can be solved as 
\begin{equation}
\label{GUP-d-4}
P_i = p_i \left(1 + \alpha {\bf p}^2 \right) + {\cal O} (\alpha^2)   \hspace{1.0cm} Q_i = q_i
\end{equation}
where $p_i$ and $q_i$ obey the usual HUP. Using Eq. (\ref{GUP-d-4}) and Feynman's path-integral technique\cite{feynman,kleinert} the Feynman propagator (or kernel) was exactly derived 
up to ${\cal O} (\alpha)$ for $D=1$ free particle case\cite{das2012,gangop2019}. Also the propagator for simple harmonic oscillator (SHO) was also derived recently in Ref. \cite{comment-1}.

In this paper we will derive the Feynman propagators for $D$-dimensional free particle and SHO systems within first order of $\alpha$ in the GUP-corrected quantum mechanics.  
In Sec. II we derive the Feynman propagators for $D=2$ free particle and SHO systems up to ${\cal O} (\alpha)$.
In Sec. III we extend the results to $D$-dimensional systems. In Sec. IV a brief conclusion is give. In appendix A we summarize several $D$-dimensional 
Gaussian integral formulas, which are frequently used in the main text.

\section{Two-dimensional Free Particle and SHO systems}

\subsection{Free particle case}

The Hamiltonian for $2$-dimensional free particle system is 
\begin{equation}
\label{free-hamil}
H_F = \frac{1}{2 m} \left( P_1^2 + P_2^2 \right) = \frac{1}{2 m} \left( p_1^2 + p_2^2 \right) + \frac{\alpha}{m} \left( p_1^2 + p_2^2 \right)^2 + {\cal O} (\alpha^2).
\end{equation}
Then, the corresponding  Schr\"{o}dinger equation can be written as 
\begin{equation}
\label{schrodinger-1}
\left[ - \frac{\hbar^2}{2 m} \left( \frac{\partial^2}{\partial q_1^2} + \frac{\partial^2}{\partial q_2^2} \right) 
+ \frac{\alpha \hbar^4}{m} \left( \frac{\partial^4}{\partial q_1^4} + \frac{\partial^4}{\partial q_2^4} + 2 \frac{\partial^2}{\partial q_1^2} \frac{\partial^2}{\partial q_2^2} \right) \right] \phi ({\bf q}) 
= {\cal E} \phi ({\bf q}).
\end{equation}
The eigenfunction and corresponding eigenvalue of the Schr\"{o}dinger equation are
\begin{equation}
\label{solution-1}
\phi ({\bm q}) = \frac{1}{2 \pi} e^{i {\bm k} \cdot {\bm q}} \hspace{2.0cm} 
{\cal E} = \frac{\hbar^2}{2 m} |{\bm k}|^2 + \frac{\alpha \hbar^4}{m} |{\bm k}|^4.
\end{equation}
Using Eqs. (\ref{solution-1}), ({\ref{app-1}), and (\ref{app-2}), one can straightforwardly derive the Feynman propagator $K_F [ {\bm q_f}, t_f: {\bm q_0}, t_0]$ as follows:
\begin{eqnarray}
\label{2dkernel-1}
&&K_F [ {\bm q_f}, t_f: {\bm q_0}, t_0]                                                                                      
 = \int_{-\infty}^{\infty} d{\rm k} \phi({\bm q_f}) \phi^* ({\bm q_0}) e^{- \frac{i}{\hbar} {\cal E} T}             \\     \nonumber
&&= \frac{m}{2 \pi i \hbar T} \left[ 1 + \frac{8 i \alpha \hbar m}{T} - \frac{8 \alpha m^2 |{\bf q_f} - {\bf q_0}|^2}{T^2} \right]            \\      \nonumber
&& \hspace{2.0cm}  \times \exp \left[ \frac{i m}{2 \hbar T} |{\bf q_f} - {\bf q_0}|^2 \left\{ 1 - 2 \alpha m^2 \left( \frac{|{\bf q_f} - {\bf q_0}|}{T} \right)^2 \right\} \right] + {\cal O} (\alpha^2)
\end{eqnarray}
where $T = t_f - t_0$. 
%
\subsection{SHO case}

For two-dimensional SHO the Hamiltonian can be written as 
\begin{equation}
\label{SHOhamil}
H_{SHO} = \frac{1}{2 m} \left( p_1^2 + p_2^2 \right) + \frac{\alpha}{m} \left(p_1^2 + p_2^2 \right)^2 + \frac{1}{2} m \omega^2 \left(q_1^2 + q_2^2 \right) + {\cal O} (\alpha^2).
\end{equation}
Then, the corresponding action functional is 
\begin{equation}
\label{SHOaction}
S[q_1, q_2] = \int_0^{T} dt L(q_1, \dot{q}_1: q_2, \dot{q}_2)
\end{equation}
where
\begin{equation}
\label{SHOlagrangian}
L(q_1, \dot{q}_1: q_2, \dot{q}_2) = \frac{m}{2} \left( \dot{q}_1^2 + \dot{q}_2^2 \right) - \alpha m^3 \left( \dot{q}_1^2 + \dot{q}_2^2 \right)^2 - \frac{1}{2} m \omega^2 \left(q_1^2 + q_2^2 \right) + {\cal O} (\alpha^2).
\end{equation}
Thus, the classical solutions obey
\begin{eqnarray}
\label{classical}
&&\ddot{q}_1 + \omega^2 q_1 - 4 \alpha m^2 \left[ (3 \dot{q}_1^2 + \dot{q}_2^2) \ddot{q}_1 + 2 \dot{q}_1 \dot{q}_2 \ddot{q}_2 \right] + {\cal O} (\alpha^2) = 0           \\    \nonumber
&&\ddot{q}_2 + \omega^2 q_2 - 4 \alpha m^2 \left[ ( \dot{q}_1^2 + 3 \dot{q}_2^2) \ddot{q}_2 + 2 \dot{q}_1 \dot{q}_2 \ddot{q}_1 \right]+ {\cal O} (\alpha^2) = 0.   
\end{eqnarray}
Up to ${\cal O} (\alpha)$ Eq. (\ref{classical}) can be solved as follows
\begin{eqnarray}
\label{solution1}
&&q_1 = A_1 \cos \omega t + B_1 \sin \omega t + \alpha \Bigg[ F_1 \cos \omega t + G_1 \sin \omega t                                                 \\     \nonumber
&&\hspace{1.0cm} + \frac{m^2 \omega^2}{8} \left[ -4 C_3 (\omega t) \cos \omega t + 4 C_1 (\omega t) \sin \omega t - C_2 \cos 3 \omega t - C_4 \sin 3 \omega t \right] \Bigg]    \\    \nonumber
&&q_2 = A_2 \cos \omega t + B_2 \sin \omega t + \alpha \Bigg[ F_2 \cos \omega t + G_2 \sin \omega t                                                 \\     \nonumber
&&\hspace{1.0cm} + \frac{m^2 \omega^2}{8} \left[ -4 \tilde{C}_3 (\omega t) \cos \omega t + 4 \tilde{C}_1 (\omega t) \sin \omega t - \tilde{C}_2 \cos 3 \omega t - \tilde{C}_4 \sin 3 \omega t \right] \Bigg]
\end{eqnarray}
where
\begin{eqnarray}
\label{solution-boso-1}
&&C_1 = -3 A_1 (A_1^2 + A_2^2 + B_1^2 ) - 2 A_2 B_1 B_2 - A_1 B_2^2                                          \\      \nonumber
&&C_2 = 3 \left[ A_1^3 - 2 A_2 B_1 B_2 + A_1 (A_2^2 - 3 B_1^2 - B_2^2)   \right]                             \\      \nonumber
&&C_3 = -3 A_1^2 B_1 - 2 A_1 A_2 B_2 - A_2^2 B_1 - 3 B_1 (B_1^2 + B_2^2)                                     \\      \nonumber
&& C_4 = 3 \left[ 3 A_1^2 B_1 + 2 A_1 A_2 B_2 - B_1 (-A_2^2 + B_1^2 + B_2^2)   \right]
\end{eqnarray}
and $\tilde{C}_j = C_j (1 \leftrightarrow 2)$ for $j = 1, 2, 3, 4$. Now, we impose the boundary conditions
\begin{eqnarray}
\label{boundary}
&&q_x (t=0) = q_{0,1}    \hspace{1.0cm}  q_y (t = 0) = q_{0, 2}                                             \\    \nonumber
&&q_x (t = T) = q_{f,1}    \hspace{1.0cm}   q_y (t = T) = q_{f,2}.
\end{eqnarray}
Then, it is straightforward to show
\begin{eqnarray}
\label{solution-boso-2}
&&A_1 = q_{0,1}   \hspace{1.0cm} A_2 = q_{0, 2}  \hspace{1.0cm}  F_1 = \frac{m^2 \omega^2}{8} C_2    \hspace{1.0cm}  F_2 = \frac{m^2 \omega^2}{8} \tilde{C}_2           \\    \nonumber
&&B_1 = \frac{1}{\sin \omega T} \left( q_{f,1} - q_{0,1} \cos \omega T \right)  \hspace{1.0cm} B_2 = \frac{1}{\sin \omega T} \left( q_{f,2} - q_{0,2} \cos \omega T \right)    \\   \nonumber
&&G_1 = \frac{m^2 \omega^2}{8 \sin \omega T} \left[ (4 \omega T C_3 - C_2) \cos \omega T - 4 \omega T C_1 \sin \omega T + C_2 \cos 3 \omega T + C_4 \sin 3 \omega T \right]    \\    \nonumber
&&G_2 = \frac{m^2 \omega^2}{8 \sin \omega T} \left[ (4 \omega T \tilde{C}_3 - \tilde{C}_2) \cos \omega T - 4 \omega T \tilde{C}_1 \sin \omega T + \tilde{C}_2 \cos 3 \omega T + \tilde{C}_4 \sin 3 \omega T \right].
\end{eqnarray}
Inserting the classical solutions into the action, one can derive the classical action in a form 
\begin{equation}
\label{classicalact}
S_{cl} = S_0 + \alpha S_1 + {\cal O} (\alpha^2)
\end{equation}
where 
\begin{eqnarray}
\label{classicalact-2}
&&S_0 = \frac{m \omega}{2 \sin \omega T} \left[ \left( |{\bm q_0}|^2 + |{\bm q_f}|^2 \right) \cos \omega T - 2 {\bm q_0} \cdot {\bm q_f} \right]                       \\    \nonumber
&&S_1 = - \frac{m^3 \omega^3}{32 \sin^4 \omega T} \Bigg[ (12 \omega T + 8 \sin 2 \omega T + \sin 4 \omega T) \left(|{\bm q_0}|^4 + |{\bm q_f}|^4 \right)               \\    \nonumber
&& \hspace{2.0cm}                                                      -4 (12 \omega T \cos \omega T + 11 \sin \omega T + 3 \sin 3 \omega T) ({\bm q_0} \cdot {\bm q_f})  \left( |{\bm q_0}|^2 + |{\bm q_f}|^2 \right)         \\   \nonumber
&& \hspace{2.0cm}                                                      + 4 (4 \omega T + 2 \omega T \cos 2 \omega T + 5 \sin 2 \omega T) \left[2 ({\bm q_0} \cdot {\bm q_f})^2 + |{\bm q_0}|^2 |{\bm q_f}|^2 \right]   \Bigg].
\end{eqnarray}
When $\omega \rightarrow 0$, $S_0$ and $S_1$ reduce to 
\begin{equation}
\label{limit-1}
\lim_{\omega \rightarrow 0} S_0 = \frac{m}{2 T} |{\bm q_f} - {\bm q_0}|^2    \hspace{1.0cm} \lim_{\omega \rightarrow 0} S_1 = -\frac{m^3}{ T^3} |{\bm q_f} - {\bm q_0}|^4,
\end{equation} 
which is consistent with exponential factor of Eq. (\ref{2dkernel-1}).

Now, let us derive the Feynman propagator (or kernel) $K[{\bm q_f}, {\bm q_0}: T]$ of this system. It is well-known that the propagator is related to the Schr\"{o}dinger equation as follows:
\begin{equation}
\label{SHOschrodinger-1}
K[{\bm q_f}, {\bm q_0}: T] = \sum_{n_1, n_2} \psi_{n_1, n_2} ({\bm q_f}) \psi_{n_1, n_2}^* ({\bm q_0}) e^{-(i / \hbar) E_{n_1, n_2} T},
\end{equation}
where $\psi_{n_1,n_2} ({\bm q})$ and $E_{n_1, n_2}$ are the eigenfunction and corresponding eigenvalue of the Schr\"{o}dinger equation derived from the Haniltonian (\ref{SHOhamil}). 
If we treat the term $\frac{\alpha}{m} (p_1^2 + p_2^2)^2$ as a small perturbation, it is possible to derive the eigenvalue
\begin{eqnarray}
\label{perturbation}
&&E_{n_1, n_2} = (n_1 + n_2 + 1) + \frac{\alpha m \hbar^2 \omega^2}{2} \left[ 3 (n_1 + n_2)^2 + 5 (n_1 + n_2) - 2 n_1 n_2 + 4 \right]        \\    \nonumber
&& \hspace{7cm}+ {\cal O} (\alpha^2).  \hspace{2.0cm} (n_1, n_2 = 0, 1, 2, \cdots)
\end{eqnarray}
Of course, it is also possible to derive the eigenfunction $\psi ({\bm q})$ within first order of $\alpha$ by applying the perturbation method. 
However, the derivation of the Feynman propagator in this way is extremely cumbersome because of the following reasons. Firstly, the unperturbed Hamiltonian generates 
the degenerate eigenspectrum. Thus, we should use the perturbation method with degeneracy, which is more complicated. Even though we derive all eigenfunctions up to first order of $\alpha$,
the right-hand side of Eq. (\ref{SHOschrodinger-1}) has too many terms, and each term involves the double summation with respect to $n_1$ and $n_2$. Thus, derivation of the Feynman propagator by making use of 
Eq. (\ref{SHOschrodinger-1}) is highly complicated and tedious. Thus, we will choose the alternative method in the following. 

In Ref. \cite{comment-1} the Feynman propagator for one-dimensional SHO is derived by making use of Eq. (\ref{SHOschrodinger-1}). The final result is expressed as 
\begin{equation}
\label{k-schrodinger-13}
K[q_f, q_0: T] = \sqrt{\frac{m \omega}{2 \pi i \hbar \sin \omega T}} \left[ 1 + \alpha f(q_f, q_0: T)  \right] e^{\frac{i}{\hbar} (S_0 + \alpha S_1)} + {\cal O} (\alpha^2),
\end{equation}
where 
\begin{eqnarray}
\label{k-schrodinger-14}
&& f(q_f, q_0: T) = \frac{3 i \hbar m \omega}{8 \sin^2 \omega T}  \left( 2 \omega T + 5 \sin \omega T \cos \omega T + \omega T \cos 2 \omega T \right)                   \\     \nonumber
&& \hspace{2.8cm} - \frac{3 m^2 \omega^2}{8 \sin^3 \omega T} \Bigg[ 2 \omega T \left\{ 3 \cos \omega T (q_0^2 + q_f^2) -  2 (2 + \cos 2 \omega T )q_0 q_f  \right\}                     \\     \nonumber
&& \hspace{5.0cm} + 10 \sin \omega T (q_0^2 + q_f^2 - 2 q_0 q_f \cos \omega T) - 6 \sin^3 \omega T (q_0^2 + q_f^2)       \Bigg].
\end{eqnarray}
Of course $S_0 + \alpha S_1$ is a classical action. We assume that the elevation of dimension does not change the Feynman propagator drastically. Thus we take an ansatz as follows:
\begin{equation}
\label{ansatz}
K[{\bm q_f}, {\bm q_0}: T] = \frac{m \omega}{2 \pi i \hbar \sin \omega T}  \left[ 1 + \alpha f({\bf q_f}, {\bf q_0}: T)  \right] e^{\frac{i}{\hbar} (S_0 + \alpha S_1)} + {\cal O} (\alpha^2)
\end{equation}
where 
\begin{eqnarray}
\label{ansatz-1}
&& f({\bf q_f}, {\bf q_0}: T)  = \beta_1 \frac{ i \hbar m \omega}{ \sin^2 \omega T}  \left( 2 \omega T + 5 \sin \omega T \cos \omega T + \omega T \cos 2 \omega T \right)                   \\     \nonumber
&& \hspace{2.8cm} + \beta_2 \frac{m^2 \omega^2}{ \sin^3 \omega T} \Bigg[ 2 \omega T \left\{ 3 \cos \omega T (|{\bm q_0}|^2 + |{\bm q_f}|^2) -  2 (2 + \cos 2 \omega T ){\bm q_0} \cdot  {\bf q_f}  \right\}                     \\     \nonumber
&& \hspace{6.0cm} + 10 \sin \omega T (|{\bm q_0}|^2 + |{\bm q_f}|^2 - 2 ({\bm q_0} \cdot  {\bm q_f}) \cos \omega T)                                                                     \\    \nonumber
&&\hspace{8.0cm} - 6 \beta_3 \sin^3 \omega T (|{\bm q_0}|^2 + |{\bm q_f}|^2)       \Bigg].
\end{eqnarray}
Of course $S_0$ and $S_1$ are given in Eq. (\ref{classicalact-2}). If Eq. (\ref{ansatz-1}) is a correct ansatz, the constants $\beta_j \hspace{.2cm} (j = 1, 2, 3)$ should be determined by making use of the integral 
equation
\begin{equation}
\label{integral-eq-1}
\int d{\bm q} K[{\bm q_f}, {\bm q}: T_2] K[{\bm q}, {\bm q_0}: T_1] = K[{\bm q_f}, {\bm q_0}: T]
\end{equation}
where $T_2 = T - T_1$. Using Eq. (\ref{ansatz}) one can show that the left-hand side of  Eq. (\ref{integral-eq-1}) reduces to 
\begin{eqnarray}
\label{integral-eq-2}
&&\int d{\bm q} K[{\bm q_f}, {\bm q}: T_2] K[{\bm q}, {\bm q_0}: T_1]                                                      \\    \nonumber
&&= \left( \frac{m \omega}{2 \pi i \hbar} \right)^2 \frac{1}{\sin \omega T_1 \sin \omega T_2} \exp \left[ \frac{i m \omega}{2 \hbar} \left(  |{\bm q_f}|^2 \cot \omega T_2 +  |{\bm q_0}|^2 \cot \omega T_1 \right) \right]    \\    \nonumber
&&  \hspace{2.0cm} \times \int d{\bm q} e^{-a |{\bm q}|^2 + 2 {\bm b} \cdot {\bm q}} \Bigg[ 1 + \alpha \left\{f({\bm q}, {\bm q_0}: T_1) + f({\bm q_f}, {\bm q}: T_2) \right\}                                                       \\    \nonumber 
&& \hspace{6.0cm}  + \frac{i \alpha}{\hbar}  \left\{S_1({\bm q}, {\bm q_0}: T_1) + S_1({\bm q_f}, {\bm q}: T_2) \right\} + {\cal O} (\alpha^2) \Bigg]
\end{eqnarray}
where 
\begin{equation}
\label{integral-boso-1}
a = - \frac{i m \omega}{2 \hbar} \frac{\sin \omega T}{\sin \omega T_1 \sin \omega T_2}  \hspace{1.0cm}
{\bm b} = - \frac{i m \omega}{2 \hbar} \frac{{\bm q_f} \sin \omega T_1 + {\bm q_0} \sin \omega T_2}{\sin \omega T_1 \sin \omega T_2}.
\end{equation}
Using the integral formula (\ref{app-2}) at $D = 2$,
one can show straightforwardly 
\begin{equation}
\label{main-1}
\int d{\bm q}  \left\{S_1({\bm q}, {\bm q_0}: T_1) + S_1({\bm q_f}, {\bm q}: T_2) \right\} e^{-a |{\bm q}|^2 + 2 {\bm b} \cdot {\bm q}} = \frac{\pi}{a} e^{|{\bm b}|^2 / a} \left[S_1({\bm q_f}, {\bm q_0}: T) + \Delta S \right]
\end{equation}
where
\begin{eqnarray}
\label{main-2}
&&\Delta S = -\frac{m^3 \omega^3}{32 \sin^4 \omega T_1} \Bigg[ (12 \omega T_1 + 8 \sin 2 \omega T_1 + \sin 4 \omega T_1) \frac{2}{a^2} \left( 1 + \frac{2 |{\bm b}|^2}{a}  \right)          \\   \nonumber
&&\hspace{5.0cm}  -8 (12 \omega T_1 \cos \omega T_1 + 11 \sin \omega T_1 + 3 \sin 3 \omega T_1)\frac{{\bm q_0} \cdot {\bm b}}{a^2}                                                                         \\   \nonumber
&&\hspace{6.0cm} + 8 (4 \omega T_1 + 2 \omega T_1 \cos 2 \omega T_1 + 5 \sin 2 \omega T_1 ) \frac{|{\bm q_0}|^2}{a}                           \Bigg]                                       \\    \nonumber
&&\hspace{5.0cm}+ \bigg(T_1 \rightarrow T_2, {\bm q_0} \rightarrow {\bm q_f} \bigg).
\end{eqnarray}
By making use of Eq. (\ref{main-1}) and (\ref{app-1}) at $D=2$,  
one can show that our ansatz (\ref{ansatz}) really satisfies the integral equation (\ref{integral-eq-1}) if
\begin{equation}
\label{main-3}
\beta_1 = 1    \hspace{1.0cm}  \beta_2 = - \frac{1}{2}    \hspace{1.0cm}  \beta_3 = 1.
\end{equation}
Inserting Eq. (\ref{main-3}) into Eq. (\ref{ansatz-1}) one can derive the Feynman propagator for $D=2$ SHO system explicitly. In the $\omega \rightarrow 0$ limit it is easy to show 
\begin{equation}
\label{flimit}
\lim_{\omega \rightarrow 0} f({\bf q_f}, {\bf q_0}: T) = \frac{8 i \hbar m}{T} - \frac{8 m^2}{T^2} |{\bm q_f} - {\bm q_0}|^2,
\end{equation}
which exactly coincides with the first order of $\alpha$ in the prefactor of free-particle Feynman propagator given in Eq. (\ref{2dkernel-1}).

In order to confirm that the Feynman propagator we derived is correct, we examine the other properties of the Feynman propagator.
First, it should be a solution of the time-dependent Schr\"{o}dinger equation. In fact, one can show straightforwardly
\begin{equation}
\label{revise-1}
\left[ i \hbar \frac{\partial}{\partial T} - H_2 \right] K[{\bm q_f}, {\bm q_0}: T] = {\cal O} (\alpha^2)
\end{equation} 
where 
\begin{equation}
\label{revise-2}
H_2 = -\frac{\hbar^2}{2 m} \left( \frac{\partial^2}{\partial q_{f,1}^2} + \frac{\partial^2}{\partial q_{f,2}^2} \right) 
+ \frac{\alpha \hbar^4}{m} \left( \frac{\partial^2}{\partial q_{f,1}^2} + \frac{\partial^2}{\partial q_{f,2}^2} \right)^2 
+ \frac{1}{2} m \omega^2 |{\bm q_f}|^2.
\end{equation}
Thus, the Feynman propagator obeys the time-dependent Schr\"{o}dinger equation within the first order of $\alpha$. 
If we impose $K[{\bm q_f}, {\bm q_0}: T] = 0$ for $T < 0$, Eq. (\ref{revise-1}) is modified in a form
\begin{equation}
\label{revise-3}
\left[ i \hbar \frac{\partial}{\partial T} - H_2 \right] K[{\bm q_f}, {\bm q_0}: T] = \xi \delta ({\bm q_f} - {\bm q_0}) \delta (T)
\end{equation}
where $\xi$ is some constant. If we integrate Eq. (\ref{revise-3}) from $0^-$ to $0^+$ in $T$, we get
\begin{equation}
\label{revise-4}
\lim_{T \rightarrow 0} K[{\bm q_f}, {\bm q_0}: T] = \lim_{T \rightarrow 0} K_F[{\bm q_f}, {\bm q_0}: T] = \frac{\xi}{i \hbar}
\delta ({\bm q_f} - {\bm q_0})
\end{equation}
where $K_F[{\bm q_f}, {\bm q_0}: T]$ is a Feynman propagator for free particle given in Eq. (\ref{2dkernel-1}). 
If we integrate Eq. (\ref{revise-4}) with respect to ${\bm q_f}$ and, use Eqs. (\ref{app-1}) and (\ref{app-2}) with $D=2$, 
one can show $\xi = i \hbar + {\cal O} (\alpha^2)$. Thus, the Feynman propagator satisfies the initial condition up to first order 
of $\alpha$ in the form
\begin{equation}
\label{revise-5}
\lim_{T \rightarrow 0}  K[{\bm q_f}, {\bm q_0}: T] = \left[1 + {\cal O} (\alpha^2)\right] \delta ({\bm q_f} - {\bm q_0}).
\end{equation}

\section{$D$-dimensional free particle and SHO systems}

\subsection{Free particle case}
From the Schr\"{o}dinger equation
\begin{equation}
\label{D-schrodinger}
\left[ -\frac{\hbar^2}{2 m} {\bm \bigtriangledown}^2 + \frac{\alpha \hbar^4}{m} {\bm \bigtriangledown}^4 \right] \phi({\bm q}) = {\cal E} \phi ({\bm q})
\end{equation}
where ${\bm \bigtriangledown}^2$ is a Laplacian, it is easy to derive 
\begin{equation}
\label{D-soultion}
\phi ({\bm q}) = \frac{1}{(2 \pi)^{D/2}} e^{i {\bm k} \cdot {\bm q}}     \hspace{1.0cm}  {\cal E} = \frac{\hbar^2}{2 m} |{\bm k}|^2 + \frac{\alpha \hbar^4}{m} |{\bm k}|^4.
\end{equation} 
Then, the Feynman propagator defined as 
\begin{equation}
\label{D-free-1}
K_F [{\bm q_f}, {\bm q_0}: T] = \int d {\bm k} \phi ({\bm q_f}) \phi^{*} ({\bm q_0}) e^{-\frac{i}{\hbar} {\cal E} T}
\end{equation}
can be computed easily using the integral formulas presented in appendix A. The final expression is 
\begin{equation}
\label{Dfree}
K_{F}[{\bm q_f}, {\bm q_0}: T] = \left(\frac{m}{2 \pi i \hbar T} \right)^{D/2} \left[ 1 + \frac{D (D + 2) i \alpha \hbar m}{T} - \frac{2 (D + 2) \alpha m^2}{T^2} |{\bm q_f} - {\bm q_0}|^2 \right] e^{\frac{i}{\hbar} S_{cl}} + {\cal O} (\alpha^2)
\end{equation}
where $S_{cl}$ is a classical action given by 
\begin{equation}
\label{Dfree-action}
S_{cl} = \frac{m}{2 T} |{\bm q_f} - {\bm q_0}|^2 \left(1 - \frac{2 \alpha m^2}{T^2} |{\bm q_f} - {\bm q_0}|^2 \right).
\end{equation}

\subsection{SHO case}
We note that the classical action for this case is also $S_{cl} = S_0 + \alpha S_1 + {\cal O} (\alpha^2)$, where $S_0$ and $S_1$ have the same expressions with Eq. (\ref{classicalact-2}).   This only difference is the fact that ${\bm q_0}$ and ${\bm q_f}$ are
$D$-dimensional vector.  
Now, we take an ansatz as follows:
\begin{equation}
\label{Dansatz}
K[{\bm q_f}, {\bm q_0}: T] = \left(\frac{m \omega}{2 \pi i \hbar \sin \omega T} \right)^{D/2} \left[ 1 + \alpha f({\bf q_f}, {\bf q_0}: T)  \right] e^{\frac{i}{\hbar} (S_0 + \alpha S_1)} + {\cal O} (\alpha^2)
\end{equation}
where  we assume that the prefactor $f({\bf q_f}, {\bf q_0}: T)$ is the same form with Eq. (\ref{ansatz-1}). Then, one can show
\begin{equation}
\label{Dmain-1}
\int d{\bm q}  \left\{S_1({\bm q}, {\bm q_0}: T_1) + S_1({\bm q_f}, {\bm q}: T_2) \right\} e^{-a |{\bm q}|^2 + 2 {\bm b} \cdot {\bm q}} = \left(\frac{\pi}{a}\right)^{D/2} e^{|{\bm b}|^2 / a} \left[S_1({\bm q_f}, {\bm q_0}: T) + \Delta S \right]
\end{equation}
where
\begin{eqnarray}
\label{Dmain-2}
&&\Delta S = -\frac{(D+2) m^3 \omega^3}{32 \sin^4 \omega T_1} \Bigg[ (12 \omega T_1 + 8 \sin 2 \omega T_1 + \sin 4 \omega T_1) \frac{1}{a^2} \left( \frac{D }{4} + \frac{ |{\bm b}|^2}{a}  \right)          \\   \nonumber
&&\hspace{5.0cm}  -2  (12 \omega T_1 \cos \omega T_1 + 11 \sin \omega T_1 + 3 \sin 3 \omega T_1) \frac{{\bm q_0} \cdot {\bm b}}{a^2}                                                                         \\   \nonumber
&&\hspace{6.0cm} + 2  (4 \omega T_1 + 2 \omega T_1 \cos 2 \omega T_1 + 5 \sin 2 \omega T_1 )  \frac{ |{\bm q_0}|^2}{ a}                           \Bigg]                                       \\    \nonumber
&&\hspace{5.0cm}+ \bigg(T_1 \rightarrow T_2, {\bm q_0} \rightarrow {\bm q_f} \bigg).
\end{eqnarray}
Using Eq. (\ref{Dmain-2}) it is straightforward to show that the Feynman propagator (\ref{Dansatz}) obeys the integral equation (\ref{integral-eq-1}) only if 
\begin{equation}
\label{Dmain-3}
\beta_1 = \frac{D (D+2)}{8}   \hspace{1.0cm}   \beta_2 = -\frac{D+2}{8}    \hspace{1.0cm}   \beta_3 = 1.
\end{equation}
Thus, inserting Eq. (\ref{Dmain-3}) into Eq. (\ref{ansatz-1}), one can derive the Feynman propagator for the $D$-dimensional SHO system up to first order of $\alpha$. 
It is easy to show that the $\omega \rightarrow 0$ limit in this case reduces to Eq. (\ref{Dfree}). As we discussed in the 
case of $D=2$, it is straightforward to show that the Feynman propagator satisfies the time-dependent Schr\"{o}dinger equation 
and initial condition up to first order of $\alpha$.

\section{conclusion}
We derive the Feynman propagators for $D$-dimensional free particle and SHO systems explicitly within the first order of $\alpha$ in the GUP-corrected non-relativistic quantum mechanics. 
From Feynman propagator one can derive the energy-dependent Green's function $\hat{G} [{\bf q_f}, {\bf q_0}: \epsilon]$ by taking a Laplace transform to the Euclidean propagator $G[{\bm q_f}, {\bm q_0}: \tau] \equiv K[{\bm q_f}, {\bm q_0}: T = -i \tau]$. 
This is useful quantity because the eigenvalue and eigenfunction of Schr\"{o}dinger equation appear as minus pole and its residue of $\hat{G} [{\bf q_f}, {\bf q_0}: \epsilon]$. Also one can derive the thermal state from the Euclidean propagator 
by letting $\tau \rightarrow 1 / (k_B T)$\cite{feynman}, where $k_B$ and $T$ are Boltzmann constant and external temperature. From Eq. (\ref{2dkernel-1}) the energy-dependent 
Green's function for two-dimensional free particle case can be derived in a form 
\begin{equation}
\label{green-1}
\hat{G}_F [{\bf q_f}, {\bf q_0}: \epsilon] = \frac{m}{\pi \hbar} \left[ \left( 1 + 8 \alpha \hbar m \epsilon \right) K_0 (z) - 2 \alpha \hbar m \epsilon z K_1 (z) \right]
\end{equation}
where $z = \sqrt{2 m \epsilon / \hbar} |{\bm q_f} - {\bm q_0}|$, and $K_{\nu} (z)$ is a modified Bessel function. For the case of two-dimensional SHO the energy-dependent Green's function can be derived if $\alpha = 0 $\cite{park97}. 
This is expressed as a winding number representation in a form 
\begin{eqnarray}
\label{green-2}
&&\hat{G} [{\bf q_f}, {\bf q_0}: \epsilon] = \frac{1}{2 \pi \omega |{\bm q_f}| |{\bm q_0}|} W_{-\frac{\epsilon}{2 \omega}, \frac{m}{2}} \left(\frac{m \omega}{\hbar} q_+^2 \right) M_{-\frac{\epsilon}{2 \omega}, \frac{m}{2}} \left(\frac{m \omega}{\hbar} q_-^2 \right)
                                                                                                                                                                       \\     \nonumber
&& \hspace{5.0cm}   \times   \sum_{j = -\infty}^{\infty} \frac{\Gamma ((1 + j + \epsilon / \omega) / 2)}{\Gamma (1 + j)} e^{ij \theta}
\end{eqnarray}
where $\theta = \cos^{-1} ({\bm q_f} \cdot {\bm q_0} / (|{\bm q_f}| |{\bm q_0}|)) $, $q_+ = \max (|{\bm q_f}|, |{\bm q_0}|)$, $q_- = \min (|{\bm q_f}|, |{\bm q_0}|)$, and $W_{\kappa, \mu} (z)$ and $M_{\kappa, \mu} (z)$ are the Whittaker's functions.
However, it seems to be highly nontrivial to derive the energy-dependent Green's function within ${\cal O} (\alpha)$ due to non-trivial expression of the prefactor $f({\bf q_f}, {\bf q_0}: T)$. 

It seems to be of interest to derive the Feynman propagators for the non-relativistic and spin-$1/2$ Aharonov-Bohm systems. As shown in Ref.\cite{hagen91,park95} the spin-$1/2$ system contains a two-dimensional $\delta$-function potential. 
The effect of two-dimensional $\delta$-function was extensively discussed in the usual quantum mechanics\cite{park95,jackiw,huang} by applying the self-adjoint extension \cite{solvable,capri} and renormalization scheme. 
Furthermore, the effect of one-dimensional $\delta$-function potential was discussed recently in the GUP-corrected quantum mechanics\cite{point-1}. By applying the method of Ref. \cite{point-1} we hope to discuss how GUP affects 
 Zeeman interaction of the spin with the magnetic field in the future.


\newpage 

\begin{appendix}{\centerline{\bf Appendix A: $D$-dimensional Gaussian Integral Formulas}}

\setcounter{equation}{0}
\renewcommand{\theequation}{A.\arabic{equation}}

In this appendix we summarize the several $D$-dimensional Gaussian integral formulas, which are used in the main text frequently.
If we assume that ${\bm x}$ and ${\bm q}$ are $D$-dimensional vectors, one can derive
 \begin{eqnarray}
\label{app-1}
&&\int d{\bm q}  e^{-a |{\bm q}|^2 + 2 {\bm b} \cdot {\bm q}} = \left(\frac{\pi}{a}\right)^{D / 2} e^{|{\bm b}|^2 / a}                        \\    \nonumber
&&\int d{\bm q} |{\bm q}|^2  e^{-a |{\bm q}|^2 + 2 {\bm b} \cdot {\bm q}} = \left(\frac{\pi}{a}\right)^{D / 2} e^{|{\bm b}|^2 / a} \frac{1}{a} \left(\frac{D}{2} + \frac{|{\bm b}|^2}{a} \right)                 \\    \nonumber
&&\int d{\bm q} \left({\bm x} \cdot {\bm q} \right) e^{-a |{\bm q}|^2 + 2 {\bm b} \cdot {\bm q}} = \left(\frac{\pi}{a}\right)^{D / 2} e^{|{\bm b}|^2 / a}  \frac{{\bm x} \cdot {\bm b}}{a}.
\end{eqnarray}
Also, one can show straightforwardly 
\begin{eqnarray}
\label{app-2}
&&\int d{\bm q} \left({\bm x} \cdot {\bm q} \right)^2 e^{-a |{\bm q}|^2 + 2 {\bm b} \cdot {\bm q}} =  \left(\frac{\pi}{a}\right)^{D / 2} e^{|{\bm b}|^2 / a} \left[ \frac{|{\bm x}|^2}{2 a} + \frac{({\bm x} \cdot {\bm b})^2}{a^2} \right]   \\    \nonumber
&&\int d{\bm q} \left({\bm x} \cdot {\bm q} \right) |{\bm q}|^2 e^{-a |{\bm q}|^2 + 2 {\bm b} \cdot {\bm q}} =  \left(\frac{\pi}{a}\right)^{D / 2} e^{|{\bm b}|^2 / a} \left(\frac{D+2}{2} + \frac{|{\bm b}|^2}{a} \right)  \frac{({\bm x} \cdot {\bm b})}{a^2} \\   \nonumber
&& \int d{\bm q} |{\bm q}|^4 e^{-a |{\bm q}|^2 + 2 {\bm b} \cdot {\bm q}} =  \left(\frac{\pi}{a}\right)^{D / 2} e^{|{\bm b}|^2 / a} \frac{1}{a^2}  \left(\frac{D (D+2)}{4}  + \frac{(D+2) |{\bm b}|^2}{a} + \frac{|{\bm b}|^4}{a^2}  \right).
\end{eqnarray}

\end{appendix}

\end{document}